\pdfoutput=1

\documentclass[aps,twocolumn,superscriptaddress,floatfix,preprintnumbers,nofootinbib]{revtex4-2}

\usepackage{amsmath, float, graphicx,amssymb,multirow,pgfplots,pgfplotstable}
\usepackage{tikz,hyperref}
\usepackage{ulem}
\usepackage{comment}
\usetikzlibrary{shapes.geometric,arrows}

\begin{document}

\begin{flushright}
MI-TH-205
\end{flushright}

\title{
GLIMPSE: Graphene-based super-Light Invisible Matter Particle SEarch
}

\author{Doojin Kim}
\email{doojin.kim@usd.edu}
\affiliation{Department of Physics, University of South Dakota, Vermillion, SD 57069, USA}
\affiliation{Mitchell Institute for Fundamental Physics and Astronomy, Department of Physics and Astronomy, Texas A\&M University, College Station, TX 77843, USA}
\author{Jong-Chul Park}
\email{jcpark@cnu.ac.kr}
\affiliation{Department of Physics and Institute of Quantum Systems, Chungnam National University, Daejeon 34134, Republic of Korea}
\author{Gil-Ho Lee}
\email{lghman@postech.ac.kr}
\affiliation{Department of Physics, Pohang University of Science and Technology, Pohang 37673, Republic of Korea}
\author{Kin Chung Fong}
\email{k.fong@northeastern.edu}
\affiliation{Raytheon BBN Technologies, Quantum Information Processing Group, Cambridge, MA 02138, USA}
\affiliation{Department of Physics, Harvard University, Cambridge, MA 02138, USA}

\begin{abstract}
We propose a new dark-matter detection strategy that will potentially enable the search for super-light dark matter $m_\chi \simeq 0.1$ keV, improving the minimum detectable mass by more than three orders of magnitude compared to ongoing experiments.
This can be achieved by intimately integrating the target material, specifically the $\pi$-bond electrons in graphene, into a Josephson junction to create a highly sensitive detector capable of detecting energy deposits from dark matter as small as $\sim 0.1$ meV.
We investigate detection prospects of pg-, ng-, and $\mu$g-scale detectors by calculating the scattering rate between dark matter and free electrons confined in two-dimensional graphene, including Pauli-blocking factors and in-medium screening effects. 
We find that the proposed detector is expected to not only serve as a complementary probe of super-light dark matter but also achieve higher experimental sensitivities than other proposed experiments, assuming zero readout noise, thanks to the extremely low threshold energy of our graphene Josephson junction sensor.
\end{abstract}

\maketitle


\newcommand{\PRE}[1]{{#1}} 
\newcommand{\ul}{\underline}
\newcommand{\del}{\partial}
\newcommand{\nbox}{{\,\lower0.9pt\vbox{\hrule \hbox{\vrule height 0.2 cm
\hskip 0.2 cm \vrule height 0.2 cm}\hrule}\,}}

\newcommand{\postscript}[2]{\setlength{\epsfxsize}{#2\hsize}
   \centerline{\epsfbox{#1}}}
\newcommand{\gweak}{g_{\text{weak}}}
\newcommand{\mweak}{m_{\text{weak}}}
\newcommand{\mplanck}{M_{\text{Pl}}}
\newcommand{\mstar}{M_{*}}
\newcommand{\sigmaan}{\sigma_{\text{an}}}
\newcommand{\sigmatot}{\sigma_{\text{tot}}}
\newcommand{\sigmaSI}{\sigma_{\rm SI}}
\newcommand{\sigmaSD}{\sigma_{\rm SD}}
\newcommand{\OmegaM}{\Omega_{\text{M}}}
\newcommand{\OmegaDM}{\Omega_{\text{DM}}}
\newcommand{\ipb}{\text{pb}^{-1}}
\newcommand{\ifb}{\text{fb}^{-1}}
\newcommand{\iab}{\text{ab}^{-1}}
\newcommand{\ev}{\text{eV}}
\newcommand{\kev}{\text{keV}}
\newcommand{\mev}{\text{MeV}}
\newcommand{\gev}{\text{GeV}}
\newcommand{\tev}{\text{TeV}}
\newcommand{\pb}{\text{pb}}
\newcommand{\mb}{\text{mb}}
\newcommand{\cm}{\text{cm}}
\newcommand{\m}{\text{m}}
\newcommand{\km}{\text{km}}
\newcommand{\kg}{\text{kg}}
\newcommand{\g}{\text{g}}
\newcommand{\s}{\text{s}}
\newcommand{\yr}{\text{yr}}
\newcommand{\Mpc}{\text{Mpc}}
\newcommand{\etal}{{\em et al.}}
\newcommand{\eg}{{\em e.g.}}
\newcommand{\ie}{{\em i.e.}}
\newcommand{\ibid}{{\em ibid.}}
\newcommand{\Eqref}[1]{Equation~(\ref{#1})}
\newcommand{\secref}[1]{Sec.~\ref{sec:#1}}
\newcommand{\secsref}[2]{Secs.~\ref{sec:#1} and \ref{sec:#2}}
\newcommand{\Secref}[1]{Section~\ref{sec:#1}}
\newcommand{\appref}[1]{App.~\ref{sec:#1}}
\newcommand{\figref}[1]{Fig.~\ref{fig:#1}}
\newcommand{\figsref}[2]{Figs.~\ref{fig:#1} and \ref{fig:#2}}
\newcommand{\Figref}[1]{Figure~\ref{fig:#1}}
\newcommand{\tableref}[1]{Table~\ref{table:#1}}
\newcommand{\tablesref}[2]{Tables~\ref{table:#1} and \ref{table:#2}}
\newcommand{\Dsle}[1]{\slash\hskip -0.28 cm #1}
\newcommand{\met}{{\Dsle E_T}}
\newcommand{\mpt}{\not{\! p_T}}
\newcommand{\Dslp}[1]{\slash\hskip -0.23 cm #1}
\newcommand{\Dsl}[1]{\slash\hskip -0.20 cm #1}

\newcommand{\mB}{m_{B^1}}
\newcommand{\mq}{m_{q^1}}
\newcommand{\mf}{m_{f^1}}
\newcommand{\mKK}{m_{KK}}
\newcommand{\WIMP}{\text{WIMP}}
\newcommand{\SWIMP}{\text{SWIMP}}
\newcommand{\NLSP}{\text{NLSP}}
\newcommand{\LSP}{\text{LSP}}
\newcommand{\mWIMP}{m_{\WIMP}}
\newcommand{\mSWIMP}{m_{\SWIMP}}
\newcommand{\mNLSP}{m_{\NLSP}}
\newcommand{\mchi}{m_{\chi}}
\newcommand{\mgravitino}{m_{\gravitino}}
\newcommand{\mmed}{M_{\text{med}}}
\newcommand{\gravitino}{\tilde{G}}
\newcommand{\Bino}{\tilde{B}}
\newcommand{\photino}{\tilde{\gamma}}
\newcommand{\stau}{\tilde{\tau}}
\newcommand{\slepton}{\tilde{l}}
\newcommand{\snu}{\tilde{\nu}}
\newcommand{\squark}{\tilde{q}}
\newcommand{\mgaugino}{M_{1/2}}
\newcommand{\epsEM}{\varepsilon_{\text{EM}}}
\newcommand{\mmess}{M_{\text{mess}}}
\newcommand{\lmess}{\Lambda}
\newcommand{\nmess}{N_{\text{m}}}
\newcommand{\signmu}{\text{sign}(\mu)}
\newcommand{\Omegachi}{\Omega_{\chi}}
\newcommand{\lambdafs}{\lambda_{\text{FS}}}
\newcommand{\be}{\begin{equation}}
\newcommand{\ee}{\end{equation}}
\newcommand{\bea}{\begin{eqnarray}}
\newcommand{\eea}{\end{eqnarray}}
\newcommand{\beq}{\begin{equation}}
\newcommand{\eeq}{\end{equation}}
\newcommand{\beqn}{\begin{eqnarray}}
\newcommand{\eeqn}{\end{eqnarray}}
\newcommand{\baln}{\begin{align}}
\newcommand{\ealn}{\end{align}}
\newcommand{\lsim}{\lower.7ex\hbox{$\;\stackrel{\textstyle<}{\sim}\;$}}
\newcommand{\gsim}{\lower.7ex\hbox{$\;\stackrel{\textstyle>}{\sim}\;$}}

\newcommand{\ssection}[1]{{\em #1.\ }}
\newcommand{\rem}[1]{\textbf{#1}}

\def\ie{{\it i.e.}\/}
\def\eg{{\it e.g.}\/}
\def\etc{{\it etc}.\/}
\def\calN{{\cal N}}

\def\mptwo{{m_{\pi^0}^2}}
\def\mp{{m_{\pi^0}}}
\def\sqtsn{\sqrt{s_n}}
\def\sqtsn{\sqrt{s_n}}
\def\sqtsn{\sqrt{s_n}}
\def\sqts0{\sqrt{s_0}}
\def\Dsqts{\Delta(\sqrt{s})}
\def\Omegatot{\Omega_{\mathrm{tot}}}

\newcommand{\changed}[2]{{\protect\color{red}\sout{#1}}{\protect\color{blue}\uwave{#2}}}


\section{Introduction}

While dark matter is a crucial ingredient of the universe and its cosmological history, the elusive nature of dark matter renders its detection via non-gravitational interactions rather challenging.
A host of theoretical and experimental efforts have been devoted to understanding weakly interacting massive particles (WIMPs), mainly motivated by the so-called WIMP miracle. 
The null signal observations thus far set stringent bounds on dark-matter candidates of $m_\chi \sim 10~{\rm GeV}-100 ~{\rm TeV}$~\cite{PandaX-II:2020oim, LZ:2022lsv, XENON:2023cxc, COSINE-100:2025kbw, IceCube:2016dgk, CMS:2017jdm, ATLAS:2017nga, PandaX-II:2016wea, LUX:2017ree}.
On the other hand, the dark-matter mass can range, in general, from $10^{-22}$ eV to $10^{68}$ eV~\cite{Battaglieri:2017aum}, so the spotlight is directed gradually toward other mass scales.

Dark-matter candidates lying in the keV-to-MeV mass range have received particular attention.
Most of the conventional dark-matter direct-detection experiments
are sensitive to energy deposits greater than $\mathcal{O}(1~{\rm eV})$, and therefore lose sensitivity to signals from dark matter lighter than $\mathcal{O}(1~{\rm MeV})$.
As a result, relevant dark-matter models remain less constrained. 
Moreover, the thermal production of such dark matter is still viable. 
Numerous models predicting dark-matter candidates within this mass range have been proposed, each addressing various interesting phenomenological aspects. 
Examples include self-/strongly-interacting dark matter~\cite{Carlson:1992fn, Hochberg:2014dra}, dark-matter freeze-in production~\cite{Moroi:1993mb, Hall:2009bx}, keV mirror-neutrino dark matter~\cite{Berezhiani:1995yi},  MeV dark matter for the 511 keV $\gamma$-ray line~\cite{Boehm:2003bt}, MeV secluded dark matter~\cite{Huh:2007zw, Pospelov:2007mp}, keV dark matter for the 3.5 keV X-ray line~\cite{Kong:2014gea, Kim:2015gka}, and elastically decoupling dark matter~\cite{Kuflik:2015isi}.

Direct searches for MeV-range light dark matter are being actively performed by experiments such as the DAMIC, DarkSide, EDELWEISS, SENSEI, SuperCDMS, and XENON1T Collaborations. 
However, experimental detection of keV-range ``super-light'' dark matter is very challenging, as the expected energy deposition is on the order of ${\rm meV}-{\rm eV}$, requiring a tiny energy threshold.
Several detection schemes have been proposed thus far~\cite{Hochberg:2015pha,Hochberg:2015fth,Hochberg:2016ntt,Cavoto:2017otc,Hochberg:2017wce,Baracchini:2018wwj,Hochberg:2019cyy,Blanco:2019lrf,Schutz:2016tid,Knapen:2016cue,Maris:2017xvi,Essig:2019kfe,Knapen:2017ekk,Griffin:2018bjn}, based on new technologies measuring small energy depositions in the detector.
They assumed that dark matter scatters off electrons~\cite{Hochberg:2015pha,Hochberg:2015fth,Hochberg:2016ntt,Cavoto:2017otc,Hochberg:2017wce,Baracchini:2018wwj,Hochberg:2019cyy,Blanco:2019lrf}, nuclei~\cite{Schutz:2016tid,Knapen:2016cue,Maris:2017xvi}, molecules~\cite{Essig:2019kfe}, or phonons~\cite{Knapen:2017ekk,Griffin:2018bjn} in the detector material. 
Various detector materials were investigated, including superconducting devices~\cite{Hochberg:2015pha,Hochberg:2015fth,Hochberg:2019cyy}, superfluid helium~\cite{Schutz:2016tid,Knapen:2016cue,Maris:2017xvi,Hertel:2018aal,vonKrosigk:2022vnf,SPICE:2023tru}, graphene~\cite{Hochberg:2016ntt,Baracchini:2018wwj}, carbon nanotube~\cite{Cavoto:2017otc}, three-dimensional Dirac materials~\cite{Hochberg:2017wce,Geilhufe:2019ndy} and polar materials~\cite{Knapen:2017ekk,Griffin:2018bjn}.\footnote{Quantum material-based detection principles, initially proposed for neutrino detection~\cite{Winslow:2012ey}, can also be effectively utilized for light dark-matter detection.}
Although some of them may accommodate dark-matter events invoking meV-range energy depositions, the actual detection is essentially limited by bolometer technology.
The aforementioned proposals utilize sensors such as a transition-edge sensor (TES)~\cite{doi:10.1063/1.1770037}, a microwave kinetic inductance device (MKID)~\cite{Day2003}, and a superconducting-nanowire single-photon detector (SNSPD)~\cite{doi:10.1063/1.1388868}, with typical operation frequencies ranging from X-ray to near-infrared~\cite{Gerrits2016}, from X-ray to far-infrared~\cite{doi:10.1146/annurev-conmatphys-020911-125022}, and from ultraviolet to mid-infrared~\cite{Holzman:2018dkq}, respectively.
To detect the energy deposition from dark matter as small as meV (i.e. $\sim240$ GHz), further R\&D is therefore needed.

To extend the search for dark matter to a lower mass regime, which has never been explored by dark-matter direct searches, we propose a new experimental strategy using a graphene-based Josephson-junction (GJJ) microwave single-photon detector~\cite{PhysRevApplied.8.024022}.
It was recently demonstrated in the laboratory that the GJJ device can have an energy resolution equivalent to sub-meV energy quanta~\cite{GJJ}. 
Therefore, it is possible to design and embark on experiments aiming at detecting dark matter of $m_\chi \gtrsim 0.1$ keV, using a sensor of a higher technology readiness level. 
We provide a conceptual detector-design proposal and study the detection prospects of super-light dark matter interacting with $\pi$-bond electrons not only at the proposed detector but at its smaller version prototype.

The remainder of this paper is structured as follows. 
We first discuss the principle of detecting super-light dark matter at GLIMPSE in Sec.~\ref{sec:principle}. 
This is followed by a conceptual design proposal in  Sec.~\ref{sec:concept}. 
The formalism for calculating the event rate is detailed in Sec.~\ref{sec:event}, highlighting the two-dimensional characteristics of graphene and their impact on the calculation. 
The sensitivity estimates expected at GLIMPSE are presented in Sec.~\ref{sec:results}, followed by a discussion of related issues and additional applications in Sec.~\ref{sec:discussions}. 
Finally, our conclusions are summarized in Sec.~\ref{sec:conclusions}.

\section{Detection principle}\label{sec:principle}

\begin{figure}[t]
\centering
\includegraphics[width=7.8cm]{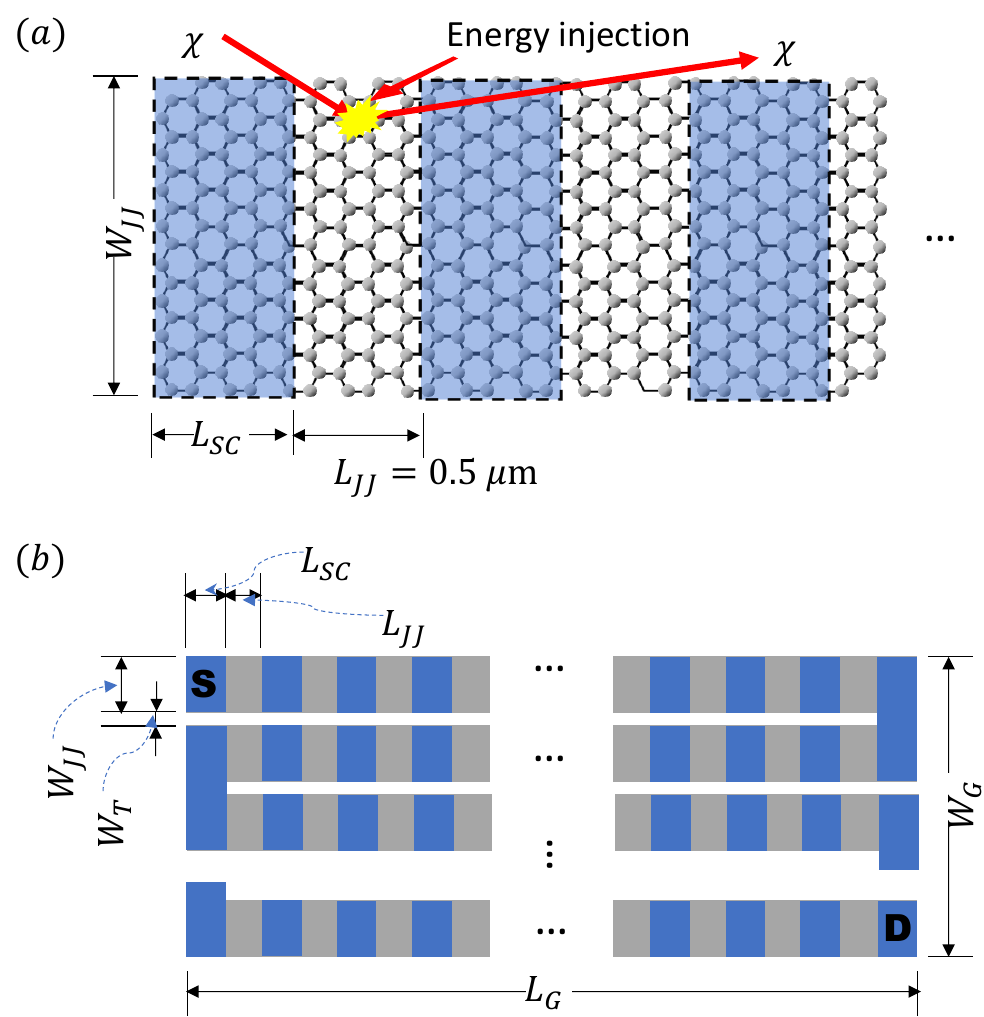}
\caption{\label{fig:principle} ($a$) Schematic description of detection principle with a single strip. 
Dark matter scattering can induce an energy injection.
($b$) Schematic layout of a full detector unit.
$W_{JJ}$ is the width of a graphene strip, $L_{SC}$ is the length of a superconducting electrode, and $L_{JJ}$ is the GJJ length. 
$W_G$ and $L_G$ are, respectively, the width and length of a graphene sheet, and $W_T$ is the width of trenches. 
S and D represent a pair of source and drain electrodes.
}
\end{figure}

A single unit of the device consists of a monolayer graphene sheet, with both sides connected to a superconducting material, forming a superconductor-normal metal-superconductor (SNS) Josephson junction (JJ)~\cite{PhysRevApplied.8.024022}, as schematically shown in Figure~\ref{fig:principle}($a$).  
Basically, when the injected energy raises the electron temperature in the graphene sheet, the calorimetric effect can switch the zero-voltage of JJ to a non-zero voltage state with an appropriate level of bias current.
Graphene is a monolayer of carbon atoms arranged in a hexagonal structure~\cite{Novoselov:2005kj, Zhang:2005zz}. 
Its electronic band structure shows a linear energy-momentum dispersion relationship similar to that of massless Dirac fermions in two dimensions. 
Near the Dirac point where the density of state vanishes, the electronic heat capacity also vanishes. 
Due to the extremely suppressed electronic heat capacity of monolayer graphene and its constricted thermal conductance to its phonons, the device is highly sensitive to small energy depositions. 
Recently, Lee {\it et al}.~\cite{GJJ} have demonstrated a microwave bolometer using GJJ, achieving noise equivalent power (NEP) at the thermodynamic limit. 
This NEP corresponds to an energy resolution capable of detecting a single 32-GHz (or equivalently, $\sim0.13$ meV) microwave photon in a single-photon detection mode.

If a dark-matter particle of interest couples to electrons, it can scatter off $\pi$-bond free electrons in the graphene sheet, transferring some fraction of its incoming kinetic energy, i.e., the energy injection shown in Figure~\ref{fig:principle}(a) takes place through dark matter scattering.
The recoiling electron heats up and rapidly thermalizes with nearby electrons through electron-electron interactions within a few picoseconds~\cite{Tielrooij2013,Brida2013}, and the JJ is triggered.   
Dark matter in the present universe floats around the Earth with the typical velocity being $\sim 10^{-3}c$. 
Therefore, a dark-matter particle with mass of $\mathcal{O}$(1 keV) carries a kinetic energy of $\mathcal{O}$(1 meV)$[\approx 1~{\rm keV}\times (10^{-3})^2]$.  
This suggests that the GJJ device has the sensitivity to detect signals induced even by sub-keV-range dark matter. 
To the best of our knowledge, no device has ever demonstrated this level of sensitivity in practice. 
Therefore, we anticipate that the microwave single-photon detector technology using GJJ will soon enable the exploration of parameter regions for super-light particle dark matter.

\section{Conceptual design proposal} \label{sec:concept}

Inspired by the GJJ device, we propose a dark-matter detector composed of an array of GJJ devices.
As detailed in the Appendix, we start with a large-scale graphene sheet, assuming a square shape with a width of $W_G$ and a length of $L_G$ for simplicity.
Narrow trenches with a width of $W_T=0.5~\mu$m can be cut to create multiple graphene strips, each with a width of $W_{JJ}$.
Thus, a single-detector unit is composed of multiple graphene strips of size $L_{JJ} \times W_{JJ}$ and superconducting-material strips of size $L_{SC} \times W_{JJ}$, along with a pair of source (``S'') and drain (``D'') electrodes. 
Here, $L_{SC}$ represents the length of the superconducting electrode, as illustrated in Figure~\ref{fig:principle}$(b)$.
Superconducting strips with a width of $W_{JJ}=3~\mu$m ($30~\mu$m), corresponding to a threshold energy of $\sim0.1$~meV (1~meV) are laid on a graphene strip at intervals of the GJJ length $L_{JJ}=0.5~\mu{\rm m}$. 
This arrangement forms an array of superconducting-graphene-superconducting-graphene-superconducting-$\cdots$ (SGSGS$\cdots$).
As the width of the superconducting strip increases, the graphene area also increases, leading to a higher heat capacity. 
Consequently, more energy is required to trigger the GJJ device. 
Each SGS sequence constitutes a single GJJ device. 
Notably, all GJJs are connected in series, meaning that even a single switched GJJ causes the series voltage measured between S and D to transition from zero to a finite value.
More technical details and fabrication feasibility are discussed in the Appendix.
A larger-scale detector can be constructed by stacking multiple such detector units.

Our sensitivity study aims at ng-scale and $\mu$g-scale detectors, while a pg-scale prototype detector will be developed first to assess the feasibility of building a multi-device detector.
To prepare for the unit forming a $\mu$g-scale graphene device, one needs a total of $\sim6~{\rm cm} \times 6~{\rm cm}$ sheet of graphene.
This can be achieved using commercial techniques, such as chemical vapor deposition~\cite{Kim2009, Bae2010}, which can produce monolayer graphene with a purity exceeding 95\%~\cite{Jang2015, GrapheneSquare,Graphenea}. 
We, therefore, expect that $\mu$g-scale detectors will be constructible in the near future.
In our phenomenological study, we assume a 100\% pure monolayer graphene for simplicity and illustrative purposes.\footnote{Nevertheless, we expect that realistic monolayer coverage will not significantly affect our main findings. 
Because the heat capacity increases in multilayer regions, even if a signal interaction occurs in such an area, the JJ would not fall into a resistive state, reporting no signal. 
In this scenario, we can effectively fiducialize the graphene area, resulting in only a minor reduction (a few percent) in the expected number of signal events.}

Since the GJJ bolometer is extremely sensitive to small changes in temperature, it is crucial to keep the system temperature low enough to suppress potential thermal backgrounds or noise.
To this end, we place the detector in the cryogenic environment by cooling the detector system down to $\sim 10$~mK using dilution refrigerators. 
Furthermore, external electromagnetic noise can be sufficiently suppressed by RF electrical filtering, mu-metal shielding, etc., which have been well developed for quantum computing and information technologies.\footnote{More precise estimates of instrumental backgrounds, such as infrared leakage from warmer stages and dark count rates, as well as the identification of experimental requirements to suppress these backgrounds and stabilize the cryogenic system, are beyond the scope of this paper. 
Therefore, we defer dedicated studies on these aspects to future work.}
An irreducible background may arise from the solar neutrinos (mostly $pp$ neutrinos) scattering off an electron and depositing a small amount of energy~\cite{Bahcall:1997eg}. 
However, considering the volume of the detector at hand, the expected number of background events is negligible
($\lsim \mathcal{O}(10^{-7})~$$\mu$g$^{-1}$year$^{-1}$)~\cite{Essig:2011nj, Hochberg:2015fth}.

We remark that in our study here, the proposed detector serves as not only the bolometer to measure the temperature change or the single-photon detector to count the number of scattering events but also the target material off which dark matter scatters.
This approach is conceptually similar to SNSPD-based experiments~\cite{Hochberg:2019cyy}, but contrasts with other proposals such as HeRALD~\cite{SPICE:2023tru} and PTOLEMY~\cite{Hochberg:2016ntt,Baracchini:2018wwj}. 
As an alternative utilization of the GJJ device technology, it can be used solely as a bolometer or a single-photon detector in other dark-matter detection proposals by coupling it to the target material such as superconductors, superfluid helium, or polar materials in Refs.~\cite{Hochberg:2015fth}, \cite{Schutz:2016tid}, and \cite{Knapen:2017ekk,Geilhufe:2019ndy}, respectively. 
In this case, energetic quasiparticles generated in the target material by dark-matter interactions are absorbed into graphene, raising its electronic temperature and subsequently triggering the GJJ.

\section{Event rate} \label{sec:event}

Our interest here is the event counts arising from the interaction between dark matter and $\pi$-bond electrons in graphene, which can behave like free electrons. 
Dark matter may also deposit energy by scattering off $\sigma$-bond electrons, potentially enhancing signal sensitivity. 
However, we conservatively restrict our analysis to the former channel to demonstrate the usefulness of the GJJ-based detector as a super-light dark-matter detector, while deferring a full analysis for a future publication.

While our calculations primarily follow the procedure formulated in Refs.~\cite{Hochberg:2015pha,Hochberg:2015fth}, we introduce modifications wherever the two-dimensional nature of graphene becomes relevant.
The total expected event count per unit mass per unit time, $n_{\rm eve}$ is given by
\begin{equation}
   n_{\rm eve}=\int dE_r dv_{\chi\parallel}\frac{d\langle n_{\rm gr}^e \sigma v_{{\rm rel}\parallel} \rangle}{dE_r}\frac{1}{a_{\rm gr}}\frac{\rho_\chi}{m_\chi}f_{\rm MB}(v_{\chi\parallel})\,,
\end{equation}
where $n_{\rm gr}^e$ is the number density of target electrons per unit area and $a_{\rm gr}=7.62\times 10^{-8}$ g$\cdot$cm$^{-2}$ is the areal density of graphene. 
$m_\chi$ is the mass of dark matter and $\rho_\chi$ is the local dark-matter energy density which is chosen to be 0.3~GeV$\cdot$cm$^{-3}$ throughout our analysis.
$f_{\rm MB}(v_{\chi\parallel})$ is the graphene-surface-parallel velocity profile of dark matter for which we take a plane-projection of a modified Maxwell-Boltzmann distribution $F_{\rm MB}$~\cite{Smith:2006ym} with root-mean-square velocity $v_0=220$~km$\cdot$s$^{-1}$ and escape velocity $v_{\rm esc}=550$~km$\cdot$s$^{-1}$ (see the Appendix for more details):
\begin{equation}
    f_{\rm MB}(v_{\chi\parallel})=\int^{\sqrt{1-(v_{\chi\parallel}/v_{\rm esc})^2}}_{-\sqrt{1-(v_{\chi\parallel}/v_{\rm esc})^2}} d\cos\theta\, \frac{1}{2\sin\theta}F_{\rm MB}\left(\frac{v_{\chi\parallel}}{\sin\theta} \right) \,, \label{eq:fMB}
\end{equation} 
where $\theta$ is the angle between the incoming dark-matter direction and the graphene-surface-normal direction. 
Finally, the velocity-averaged event rate on a (sufficiently thin) graphene sheet per unit time $\langle n_{\rm gr}^e \sigma v_{{\rm rel}\parallel} \rangle$ is
\begin{equation}
    \langle n_{\rm gr}^e \sigma v_{{\rm rel}\parallel} \rangle =\int  \frac{d^3p_{\chi,f}}{(2\pi)^3}  
    \frac{\overline{|\mathcal{M}|}^2}{16m_e^2m_\chi^2}\left(\frac{f_e}{f_e^0} \right)^2 S_{\rm gr}(E_r,q)\,, \label{eq:velavesig}
\end{equation}
where $p_{\chi,f}$ is the momentum of final-state dark matter and all the total energy quantities are taken in the non-relativistic limit. 
Here $\overline{|\mathcal{M}|}^2$ denotes the matrix element for the scattering process between dark matter and free electrons.
The in-medium screening effect is parameterized by the factor $f_e/f_e^0=(\hat{\bf q}\cdot {\bf \epsilon}\cdot \hat{\bf q})^{-1}$ with $\epsilon$ being the dielectric tensor of graphene. 
We follow the computational approach outlined in Ref.~\cite{Griffin:2021znd}, using the graphene dielectric function as obtained under the random phase approximation in Ref.~\cite{PhysRevB.75.205418}. 
The Pauli-blocking effects are encoded in $S_{\rm gr}(E_r,q)$, a structure-function over electron-recoil kinetic energy $E_r$ and the magnitude of momentum transfer along the graphene surface
$q=|\Vec{p}_{\chi\parallel,i}-\Vec{p}_{\chi\parallel,f}|$ 
with the subscript $i$ implying the initial state. 
It is then convenient to convert $d^3p_{\chi,f}$ to $dE_r$ and $dq$:
\begin{equation}
    \frac{d^3p_{\chi,f}}{(2\pi)^3}\, \longrightarrow\, \frac{dE_r dq}{(2\pi)^2} \frac{ 2q(E_{\chi,i}-E_r)}{\tilde{\lambda}(q^2,p_{\chi,i}^2,p_{\chi,f}^2)}\,,
\end{equation}
where $\tilde{\lambda}(x,y,z)=\sqrt{2(xy+yz+zx)-x^2-y^2-z^2}$ and we integrate out $p_{\chi,f}^z$ by pulling out the delta-function factor in $S_{\rm gr}=(2\pi)\delta(p_{\chi,i}^z-p_{\chi,f}^z)\cdot S$ (see the Appendix).
We next take the analytic expression for $S$ derived in Ref.~\cite{Hochberg:2015fth} based on Ref.~\cite{Reddy:1997yr}:
\begin{equation}
    S(E_r,q)=\frac{m_e^2T}{\pi q}\left[ \frac{E_r/T}{1-\exp(-E_r/T)}\left(1+\frac{\xi}{E_r/T} \right) \right], \label{eq:Sexp}
\end{equation}
where $T$ is the system temperature surrounding the detector which we take to be 10~mK as mentioned earlier.
The $\xi$ quantity is given by
\begin{equation}
    \xi =\log\left[ \frac{1+e^{(\epsilon_--\mu)/T}}{1+e^{(\epsilon_-+E_r-\mu)/T}} \right]
\end{equation}
with 
\begin{equation}
    \epsilon_-=\frac{1}{4}\frac{\lbrace E_r-q^2/(2m_e)\rbrace^2}{q^2/(2m_e)}\,.
\end{equation}
Here $\mu$ is the chemical potential which is identified as the Fermi energy $E_F$ at zero temperature. 
For a two-dimensional object like graphene, the linear energy-momentum dispersion suggests $E_F=v_F\sqrt{\pi n_c}$.
Here the Fermi velocity of graphene $v_F$ is $1.15\times 10^8$~cm$\cdot$s$^{-1}$~\cite{PhysRevLett.108.116404} and the carrier density $n_c$ is chosen to be $10^{12}$~cm$^{-2}$~\cite{GJJ}.

\section{Results} \label{sec:results}

\begin{figure*}[t]
    \centering
    \includegraphics[width=8.2cm]{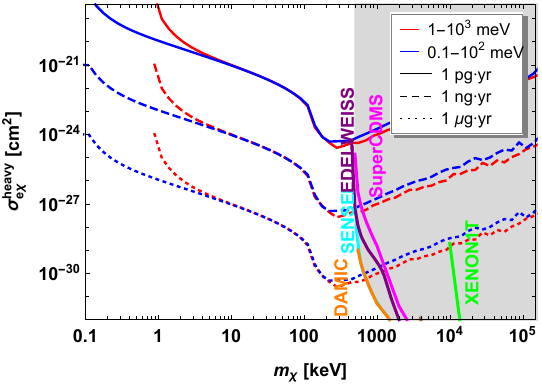} \hspace{0.5cm}
    \includegraphics[width=8.2cm]{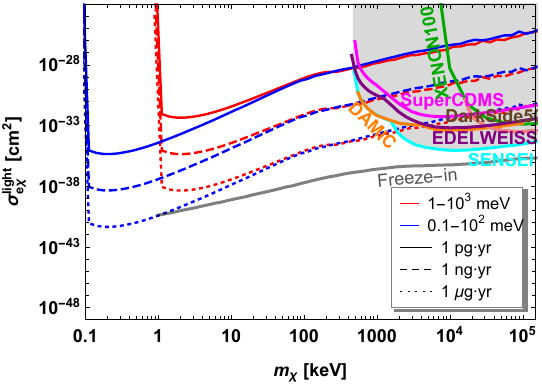}
    \caption{Sensitivities of the proposed GJJ detector to $0.1~{\rm keV}-0.1~{\rm GeV}$ dark matter in the plane of dark-matter mass $m_\chi$ and scattering cross section between dark matter and electrons $\sigma_{e\chi}$. 
    The dark matter is assumed to scatter off $\pi$-bond electrons in graphene, which can behave like free electrons, via an exchange of a heavy mediator (left panel) or a light mediator (right panel). 
    Blue (red) curves show the sensitivities with electron kinetic energy deposition falling in $0.1~{\rm meV}-0.1~{\rm eV}$ ($1~{\rm meV}-1~{\rm eV}$), and solid, dashed, and dotted curves are for 1~pg, 1~ng, and 1~$\mu$g GJJ detectors, respectively, with a year exposure.
    For comparison, we show the relic abundance line explained by freeze-in~\cite{Essig:2011nj} and the existing limits~\cite{DMLimits} (gray region) from DAMIC, DarkSide50, EDELWEISS, SENSEI, SuperCDMS, XENON100, and XENON1T.}
    \label{fig:sensitivity}
\end{figure*}

We are now ready to study the dark matter signal sensitivities achievable with the proposed GJJ detector. Following many previous studies, we assume that dark matter interacts with electrons via the exchange of a mediator $\phi$ with mass $m_\phi$.
In the non-relativistic limit, the scattering cross section between free electrons and say, fermionic dark matter $\chi$ is given by
\begin{equation}
    \sigma_{e\chi}\approx \frac{g_e^2g_\chi^2}{\pi}\frac{\mu_{e\chi}^2}{(m_\phi^2+q^2)^2}\,, \label{eq:scat}
\end{equation}
where $g_e$ and $g_\chi$ parameterize couplings of $\phi$ to electrons and $\chi$ and where $\mu_{e\chi}$ is the reduced mass of the electron-$\chi$ system. 
If the mediator is heavy enough such that $m_\phi^2 \gg q^2$, Eq.~\eqref{eq:scat} is simplified to
\begin{equation}
    \sigma_{e\chi}^{\rm heavy} \approx \frac{g_e^2g_\chi^2}{\pi}\frac{\mu_{e\chi}^2}{m_\phi^4}\,.\label{eq:heavymed}
\end{equation}
On the other hand, for the opposite limit or the light-mediator case (i.e., $m_\phi^2 \ll q^2$), we get
\begin{equation}
    \sigma_{e\chi}^{\rm light} \approx \frac{g_e^2g_\chi^2}{\pi}\frac{\mu_{e\chi}^2}{q^4}\,.\label{eq:lightmed}
\end{equation}
The matrix element in Eq.~\eqref{eq:velavesig} is related to the scattering cross section in Eq.~\eqref{eq:scat} as
\begin{equation}
    \sigma_{e\chi}=\frac{1}{16\pi }\frac{\overline{|\mathcal{M}|}^2}{m_e^2m_\chi^2} \mu_{e\chi}^2\,.
\end{equation}
To estimate our sensitivity reach, we take Eqs.~\eqref{eq:heavymed} and \eqref{eq:lightmed} as reference cross sections.
For the light-mediator case, we follow the prescription in Refs.~\cite{Essig:2011nj, Hochberg:2019cyy} and replace the $q$-dependence by a reference value $q_{\rm ref}=\alpha_e m_e$ with $\alpha_e$ being the usual electromagnetic fine-structure constant. 

Figure~\ref{fig:sensitivity} illustrates the sensitivities of the proposed GJJ detector to dark matter in the $0.1~{\rm keV}-0.1~{\rm GeV}$ range, shown in the plane of dark-mater mass $m_\chi$ and scattering cross section $\sigma_{e\chi}$, assuming a 100\% duty factor for illustration.\footnote{A precise estimate of realistic duty factors depends on factors such as re-calibration and maintenance; however, our main physics conclusions remain unchanged.}
Here we require 3.6 signal events, corresponding to the 95\% C.L. upper limit under the assumption of a null event observation, with negligible neutrino-induced background, following Poisson statistics~\cite{Feldman:1997qc}.
As mentioned before, such dark matter is assumed to interact with the $\pi$-bond electrons in the graphene target through an exchange of $\phi$. 
The left and the right panels are for the heavy- and the light-mediator cases, respectively. 
In both panels, the blue (red) curves show the sensitivities with electron kinetic-energy deposition being in-between 0.1~meV and 0.1~eV (1~meV and 1~eV), i.e., the superconducting strips of the GJJ detector have $\ell=3~\mu$m ($\ell=30~\mu$m).
The solid, dashed, and dotted curves are the expected results with 1~pg-scale, 1~ng-scale, and 1~$\mu$g-scale detectors exposed for a year, respectively. 
The pg-scale detector can be placed on the ground without being affected by non-instrumental backgrounds. 
However, the attenuation effect from the Earth's atmosphere~\cite{Emken:2019tni}, equivalent to approximately 10 m of water, would pose a challenge in achieving the expected sensitivity for the heavy mediator case (solid curves in the left panel of Figure~\ref{fig:sensitivity}.
For reference purposes, we also exhibit the relic abundance line explained by freeze-in~\cite{Essig:2011nj} and the existing bounds~\cite{DMLimits} together from DAMIC, DarkSide50, EDELWEISS, SENSEI, SuperCDMS, XENON100, and XENON1T by the orange, brown, purple, cyan, pink, dark green, and green lines, respectively.

\section{Discussions} \label{sec:discussions}

We remark that cosmological and astrophysical observations provide bounds on super-light dark matter while the aforementioned dark-matter direct-search experiments are exploring $m_\chi-\sigma_{e\chi}$ space. 
For example, the so-called Lyman-$\alpha$ forest is a powerful tool to constrain keV-range dark matter which was thermally produced in the early universe. 
Such ``warm'' dark matter appears relativistic when it freezes out, so it may affect the structure formation, leaving appreciable differences in the Lyman-$\alpha$ absorption features from what is observed today (see, e.g., Ref.~\cite{Boyarsky:2008xj}). 
Currently, warm dark matter of $m_\chi \lsim \mathcal{O}({\rm keV})$ is disfavored by Lyman-$\alpha$ forests, while there still exist relatively large uncertainties even among very recent results: e.g., Ref.~\cite{Palanque-Delabrouille:2019iyz} claims that $m_\chi>10$~keV is allowed at 95\% C.L., whereas Ref.~\cite{Garzilli:2019qki} claims $m_\chi>1.9$~keV is allowed at 95\% C.L.. 
In fact, the proposed GJJ detector can efficiently probe $1-10$~keV dark matter very efficiently by lowering the threshold to 0.1~meV, offering a significant advantage over dark matter detectors based on other bolometer technologies.
This is because a lower threshold enables access to the phase space associated with recoil electrons carrying lower energy. 
This is also clearly demonstrated by the red (higher threshold) and the blue (lower threshold) curves in Figure~\ref{fig:sensitivity}.

By contrast, if dark matter is produced non-thermally, the bounds from the Lyman-$\alpha$ forest are generically irrelevant.
It is therefore important to explore the $\mathcal{O}({\rm keV})$ mass region in a model-independent sense.
There are many well-motivated physics models containing such non-thermal super-light dark-matter candidates (see  Ref.~\cite{Baer:2014eja} for a recent review): for example, sterile neutrinos, super-light dark gauge bosons, axion-like dark-matter particles, and axino/gravitino dark matter. 
We expect that search strategies using the proposed GJJ detector can provide a probe complementary to the existing experimental effort for such dark-matter candidates.

We emphasize the implications of our simplified model, which describes the scattering between dark matter moving in three-dimensional space and free electrons confined within the two-dimensional graphene sheet.
As stated earlier, the momentum transfer $q$ in Eq.~\eqref{eq:velavesig} is mainly determined by the change of the $\chi$-momentum component parallel to the graphene surface. Therefore, the device sensitivity to the dark-matter signal will be maximized (minimized) when $\chi$ is incident in the direction parallel (perpendicular) to the graphene surface.
In other words, the dark matter signal could be validated by actively rotating the graphene sheet to align it either parallel or perpendicular to the overall dark matter flux and observing the directional dependence of event rates.\footnote{See Ref.~\cite{Pedersen:2019gdh} pointing out the potential of a 3-dimensional dark-matter detector utilizing the direction-sensitive scintillation yields of ZnWO$_4$ crystals.}
Moreover, such directional dependence of signal rates can be utilized to determine the mass scale of dark matter, which can be generally applied to the (at least effectively) two-dimensional direct detection experiments~\cite{Angle}.
We reserve further investigations on this interesting possibility and detailed subleading effects beyond our simplified scattering model for a future publication.

Finally, we point out that the proposed detector is capable of detecting even lighter dark matter candidates of sub-${\rm meV}-{\rm eV}$ mass. 
For example, axion-like particle or dark gauge boson dark matter candidates within such a mass range can be absorbed to the detector material via a Compton-like process with an electron resulting in an emission of a photon, i.e., $\chi+e\to \gamma + e$.
Here the radiated photon is typically as energetic as the mass-energy of the incoming dark matter, so the proposed detector may have decent sensitivities to these dark matter candidates at masses, and thus photon energies, below about 0.1~eV~\cite{Chang2016}, where the intraband absorption is efficient.
Note that if the mass of the absorbed dark matter particle is greater than the binding energy of an electron, i.e., $\mathcal{O}$(eV), an electron is ejected via a process analogous to the photoelectric effect~\cite{Dimopoulos:1985tm, Avignone:1986vm, Pospelov:2008jk, Bloch:2016sjj}. 
Numerous dark-matter direct detection experiments have conducted searches for such ejected electron signals, and our proposal can offer a new avenue for this search effort.

\section{Conclusions} \label{sec:conclusions}

In conclusion, we have proposed a class of {\it new} dark matter detectors, adopting the GJJ device which has been implemented and demonstrated experimentally. 
Owing to its outstanding sensitivity to energy changes as small as $\sim 0.1$ meV, the proposed detectors---comprising an array of GJJ devices---are expected, for the first time, to be capable of probing dark matter candidates as light as $\sim 0.1$ keV through dark matter-electron scattering.
We have shown that the sensitivity of detectors of 1-$\mu$g graphene could reach $\sigma_{e\chi}\approx 10^{-31}~{\rm cm}^{-2}$ and $\sigma_{e\chi}\approx 10^{-42}~{\rm cm}^{-2}$ for the heavy and the light mediator cases with one-year exposure. 
To evaluate the maximum potential of the proposed concept, we estimated the expected dark-matter signal sensitivity under the assumption of zero readout noise.

Finally, we emphasize that the primary focus of our work is to highlight the phenomenological potential of a novel dark-matter detection concept enabled by a new device technology while deferring dedicated studies on potential experimental and practical challenges, as well as the requirements to address them, to follow-up work. 
In this direction, we are currently developing a prototype detector with a larger SGS JJ multiplicity, along with more detailed background studies and extended investigations into various dark-matter candidates that could be detected with the proposed detectors.

\bigskip

\noindent{{\it Note added}. During the review process, we became aware of light DM direct detection studies utilizing other technologies, e.g. sub-GeV nuclear recoil techniques~\cite{Blanco:2022pkt, Romao:2023zqf}, graphene-based detection~\cite{Catena:2023qkj, Catena:2023awl},  detection techniques involving $\pi$-electron transitions~\cite{Blanco:2021hlm}, and quantum material-based detection techniques~\cite{Blanco:2022cel}.}


\section*{Acknowledgments}

We would like to thank Kaustubh Agashe, Bhaskar Dutta, and Yue Zhao for their careful reading of the draft and insightful discussions.
We also would like to thank Yonit Hochberg,  Rupak Mahapatra, and Hongki Min for their useful comments and discussions.  
This work was performed in part at the Aspen Center for Physics, which is supported by National Science Foundation grant PHY-1607611.
Part of this work was discussed in the workshop ``Dark Matter as a Portal to New Physics'' at the Asia Pacific Center for Theoretical Physics.
The work of DK is supported by the U.S. Department of Energy Grant DE-SC0010813.
JCP and GHL acknowledge the support of the Samsung Science and Technology Foundation (Project No. SSTF-BA2101-06).
KCF acknowledges the support from the Army Research Office under Cooperative Agreement Number W911NF-17-1-0574.

\section*{Appendix}

\begin{figure*}[t]
    \centering
    \includegraphics[width=15.5cm]{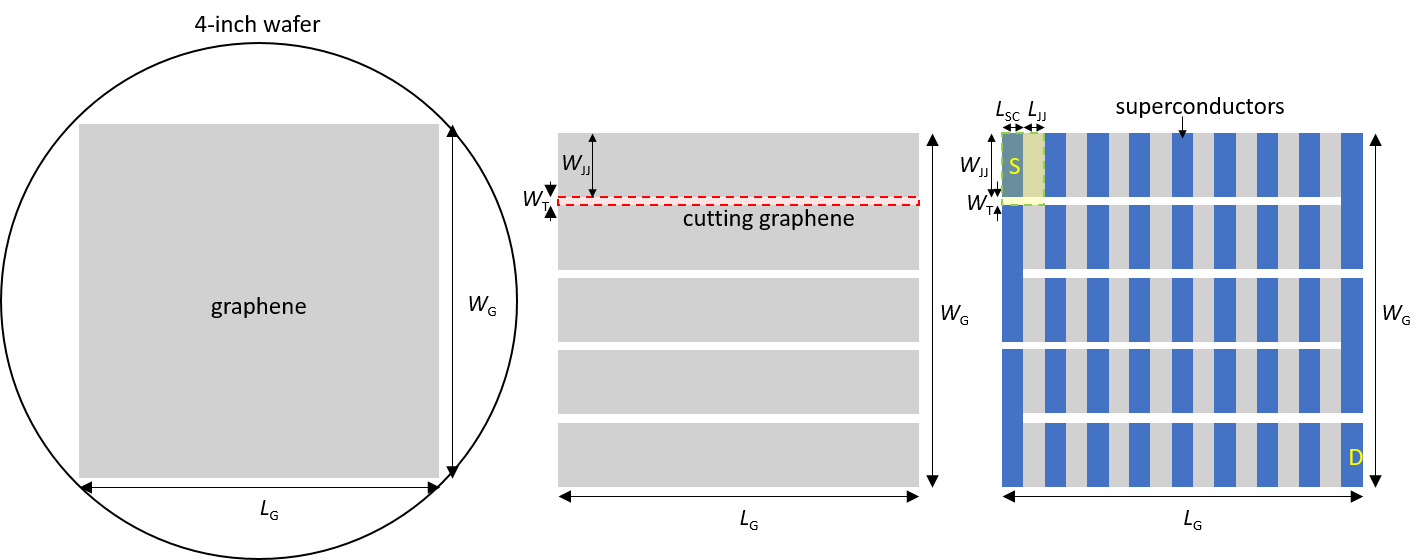} 
    \caption{Schematic representation of a possible fabrication procedure. The schematics are not to scale.
    ($a$) A graphene sheet of $W_G\times L_G$ grown on a 4-inch wafer.
    ($b$) The etched graphene represented by a box outlined with the dotted red lines.
    ($c$) A superconductor-loaded full detector unit.
    } \label{fig:fab}
\end{figure*}

\noindent {\bf \textit{Detector design details.}}
We provide more technical details for building a large-scale detector consisting of numerous GJJ units.
We show a possible architecture for making $\sim 10^{9}$ GJJs, the number that would be required for a $\mu$g-scale detector.
As shown in Figure~\ref{fig:fab}, we begin with a wafer-scale graphene\footnote{Chemical vapor deposition grown monolayer graphene of $1~{\rm m}\times 1~{\rm m}$ is already available in the market such as Graphene Square~\cite{GrapheneSquare}. 
Even high-purity ($>95\%$) monolayer graphenes can be grown by companies such as Graphene Square~\cite{GrapheneSquare} (8-in wafer) and Graphenea~\cite{Graphenea} (4-in wafer).} grown on a 4-inch silicon wafer covered by hexagonal boron nitride insulating layers and assume that the graphene sheet has a square shape with a width $W_G=6$~cm  and a length $L_G=6$~cm, for simplicity in our calculation.
Narrow trenches with a width $W_T=0.5~\mu$m can be cut to create multiple strips of graphene, each with a width of $W_{JJ}=3~\mu$m.
The superconducting electrode can be deposited in such a way that all the GJJs are connected in series using the superconducting electrode with a length $L_{SC}=0.5~\mu$m and a GJJ length $L_{JJ}=0.5~\mu$m.
As the smallest feature would be 500~nm, either electron beam lithography (for example, the ELS-F150 model from Elionix can make 4 nm features and support an 8-inch wafer) or photo-lithography (50 nm features can be made with deep ultraviolet light) can be used.

As mentioned in the main manuscript, it is important to note that all GJJs are connected in series, not in parallel. This configuration ensures that even if only one GJJ switches on, the series resistance measured between the source (``S'') and drain (``D'') electrodes will transition from zero to a finite value.
More specifically, we bias a DC current and monitor the DC voltage between the S electrode (i.e., the very first superconducting electrode) and the D electrode (i.e., the very last superconducting electrode).
Due to the self-heating effect in the resistive state, a GJJ will remain in a resistive state after it switches from a supercurrent state to a resistive state.
The electron cooling time in GJJ is, in principle, expected to be $\mathcal{O}(\mu{\rm s}) - \mathcal{O}({\rm ns})$~\cite{PhysRevApplied.8.024022}; therefore, the dead time could ideally be reduced to $\mathcal{O}({\rm ns})$. 
Implementing such a rapid reset is unnecessary at this stage, as we focus on detecting rare dark-matter events. 
Thus, a dead time of $\mathcal{O}({\rm ms})$ is sufficient for our purposes.
If any one of the GJJs switches, the DC voltage will spike up to some finite value since all the GJJs are connected in series, and this voltage can be easily detected even with a slow DC measurement.

The total number of GJJs on a single wafer $N_W$ is 
\begin{equation}
N_W=\frac{W_G L_G}{(W_T+W_{JJ})(L_{SC}+L_{JJ})}\approx 1.03 \times 10^9\,,
\end{equation}
which is comparable to or less than the number of transistors in modern integrated circuit chips ($\sim5-10 \times 10^9$) and the structure of GJJ is much simpler than the transistors with gates.
We therefore expect that the proposed device can be realized using currently available nanofabrication technology.
One example device consisting of multiple Josephson junctions is the voltage standard: about 20,000 Josephson junctions were fabricated for realizing a 12 V voltage standard~\cite{192296}. 

Large enough dilution refrigerator to cool down the whole device down to 10~mK temperature is already available in the market.
For example, the Proteox model from Oxford Instruments has a mixing chamber plate of diameter of 360 mm, the XLD model from Bluefors has a mixing chamber plate of diameter of 500 mm, and the CF-CS110 model from Leiden Cryogenics has a mixing chamber plate of diameter of 490 mm.

\bigskip
\noindent {\bf \textit{Projected Maxwell-Boltzmann distribution.}}
When a dark matter particle of velocity $v_\chi$ is incident on a graphene sheet by angle $\theta$ with respect to the plane-normal direction, the parallel component, i.e., $v_{\chi\parallel}=v_\chi \sin\theta$, is involved in the momentum transfer. 
Denoting the probability density in $v_{\chi\parallel}$ by $f_{\rm MB}(v_{\chi\parallel})=\frac{dP}{dv_{\chi\parallel}}$, we find that
\begin{equation}
    \frac{dP}{dv_{\chi\parallel}} =  \int \frac{dv_\chi}{dv_{\chi\parallel}} \frac{d^2P}{dv_\chi d\Omega}d\Omega =  \int \frac{1}{\sin\theta} \frac{d^2P}{dv_\chi d\Omega}d\Omega\,,
\end{equation}
where $\frac{dP}{dv_\chi}$ is a modified Maxwell-Boltzmann distribution $F_{\rm MB}$~\cite{Smith:2006ym} and $\frac{d^2P}{dv_\chi d\Omega}$ is the distribution per unit solid angle:
\begin{equation}
    \frac{F_{\rm MB}(v_\chi)}{4\pi} = \frac{v_\chi^2 \exp\left(-\frac{v_\chi^2}{v_0^2} \right)}{\pi^{3/2}v_0^3\left[{\rm erf}\left(\frac{v_{\rm esc}}{v_0}\right)-\frac{2}{\sqrt{\pi}}\frac{v_{\rm esc}}{v_0}\exp\left(-\frac{v_{\rm esc}^2}{v_0^2} \right) \right]}\,,
\end{equation}
To evaluate $f_{\rm MB}$ at $v_{\chi\parallel}$, we have to integrate over all relevant $\theta$ and azimuth $\varphi$ values:
\begin{equation}
    f_{\rm MB}(v_{\chi\parallel})=\int_0^{2\pi} d\varphi\int^{\cos\theta_+}_{\cos\theta_-} d\cos\theta\, \frac{1}{\sin\theta}\frac{F_{\rm MB}\left(\frac{v_{\chi\parallel}}{\sin\theta} \right)}{4\pi} \,,
\end{equation} 
where $\cos\theta_{\pm} = \pm\sqrt{1-\left(\frac{v_{\chi\parallel}}{v_{\rm esc}}\right)^2}$.

\begin{figure}[t]
    \centering
    \includegraphics[width=0.45\textwidth]{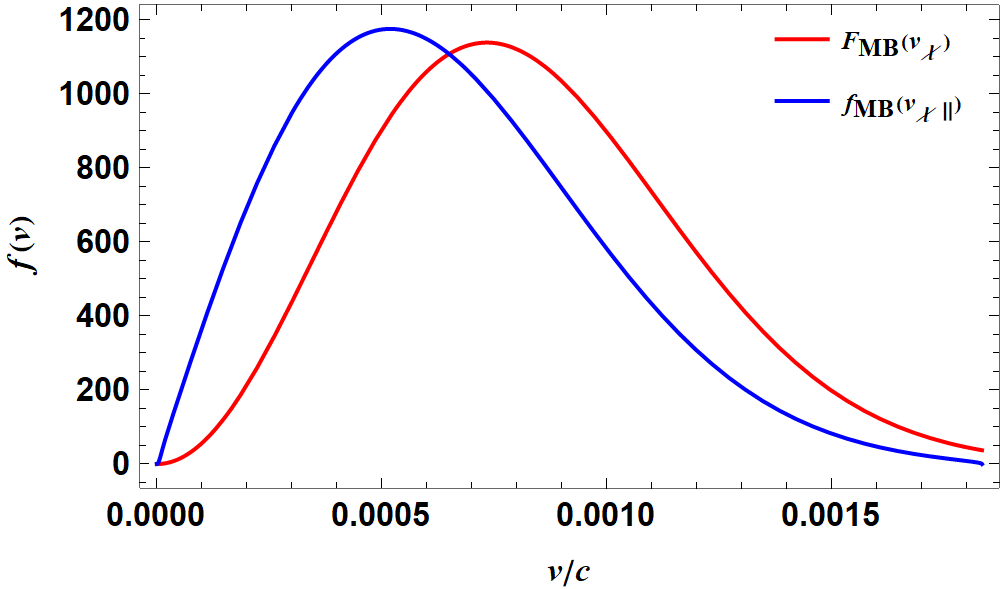}
    \caption{Comparison between $F_{\rm MB}(v_\chi)$ and $f_{\rm MB}(v_{\chi\parallel})$.}
    \label{fig:fMBcompare}
\end{figure}

We show a comparison between our projected distribution $f_{\rm MB}(v_{\chi\parallel})$ (blue curve) and the original distribution $F_{\rm MB}(v_\chi)$ (red curve) in Figure~\ref{fig:fMBcompare}. 
As expected, the graphene plane-parallel component $v_{\chi\parallel}$ populates more in the lower velocity region.

\bigskip
\noindent {\bf \textit{Structure function.}} 
The structure-function for the graphene system of interest $S_{\rm gr}$ is given by
\begin{eqnarray}
    S_{\rm gr}(E_r,q)&=&2 \int \frac{d^3p_{e,i}}{(2\pi)^3}\int \frac{d^3p_{e,f}}{(2\pi)^3}(2\pi)\delta(p_{e,i}^z-p_{e,f}^z) \nonumber \\
    &\times& (2\pi)^4\delta^{(4)}(p_{\chi,i}+p_{e,i}-p_{\chi,f}-p_{e,f}) \nonumber \\
    &\times& f_{e,i}(E_{e,i})\lbrace 1-f_{e,f}(E_{e,f}) \rbrace\,,
\end{eqnarray}
where $p_{e,i(f)}$ denotes the momentum of the initial-state (final-state) electron.
The delta function in the first line reflects the assumption that the free electrons are confined in the graphene surface, or equivalently, they do not get any significant momentum change along the surface-normal direction, as far as the electron recoil kinetic energy is sufficiently smaller than the work function of graphene. 
The Fermi-Dirac distribution functions for the initial-state (final-state) electrons $f_{e,i(f)}$ are
\begin{equation}
    f_{e,i(f)}=\left\{1+\exp\left(\frac{E_{e,i(f)}-\mu}{T} \right) \right\}^{-1}\,,
\end{equation}
where $\mu$ and $T$ are the chemical potential and the system temperature, respectively.
Integrating over $d^3p_{e,f}$ in combination with the spatial components of the four-dimensional delta function yields
\begin{eqnarray}
    S_{\rm gr}(E_r,&&q)=(2\pi)\delta(p_{\chi,i}^z-p_{\chi,f}^z)\cdot \frac{1}{2\pi^2}\int d^3p_{e,i} \nonumber \\
    &&\times \delta(E_r+E_{\chi,i}-E_{\chi,f})f_{e,i}(E_{e,i})\lbrace 1-f_{e,f}(E_{e,f}) \rbrace \nonumber \\ 
    &&\equiv (2\pi)\delta(p_{\chi,i}^z-p_{\chi,f}^z)\cdot S(E_r,q)\,,
\end{eqnarray}
where we factor out the delta function of $p_{\chi,f}^z$, which is used for the $d^3p_{\chi,f}$ integral, and separately define $S(E_r,q)$. 
The closed form for $S(E_r,q)$ is available in the non-relativistic limit~\cite{Reddy:1997yr}, as shown in Eq.~\eqref{eq:Sexp} of the main text.

\bibliographystyle{apsrev4-1}
\bibliography{ref}

\end{document}